\documentclass[pra,showpacs,twocolumn]{revtex4}%
\input{psfig.sty}
\usepackage{amsmath}
\usepackage{graphicx}
\usepackage{amsfonts}
\usepackage{amssymb}
\usepackage{ulem}%
\usepackage{color}

\begin{document}
\title{Entanglement of a photon and a collective atomic excitation}
\date{\today }
\author{D. N. Matsukevich, T. Chaneli\`{e}re, M. Bhattacharya, S.-Y. Lan, S. D. Jenkins, T.A.B. Kennedy, and A. Kuzmich}
\affiliation{School of Physics, Georgia Institute of Technology, Atlanta, Georgia 30332-0430} 
\begin{abstract}
We describe a new experimental approach to probabilistic atom-photon (signal) entanglement. Two qubit states are encoded as orthogonal collective
spin excitations of an unpolarized atomic ensemble. After a programmable delay, the atomic excitation is converted into a photon (idler).
Polarization states of both the signal and the idler are recorded and are found to be in violation of the Bell inequality. Atomic coherence times
exceeding several microseconds are achieved by switching off all the trapping fields - including the quadrupole magnetic field of the
magneto-optical trap - and zeroing out the residual ambient magnetic field.
\end{abstract}
\maketitle

Long-distance quantum cryptographic key distribution (QCKD) is an important goal of quantum information science. Extending the reach of quantum
cryptography ideally involves the ability to entangle two distant qubits (two level quantum systems) \cite{chuang,briegel}, using the Bell
inequality violation to verify the security of the quantum communication channel.  Parametric down conversion is an established technology
producing entangled photon pairs. Unfortunately, it is not directly applicable to long-distance QCKD, as the rate scales exponentially with the
distance due to probabilistic nature of entangled photon pairs generation. It is necessary to provide a controllable delay between the two
photons, that is, to have a means of photon storage.

\begin{figure}[htp]
\vspace{-0.3cm}
\begin{center}
\leavevmode  \psfig{file=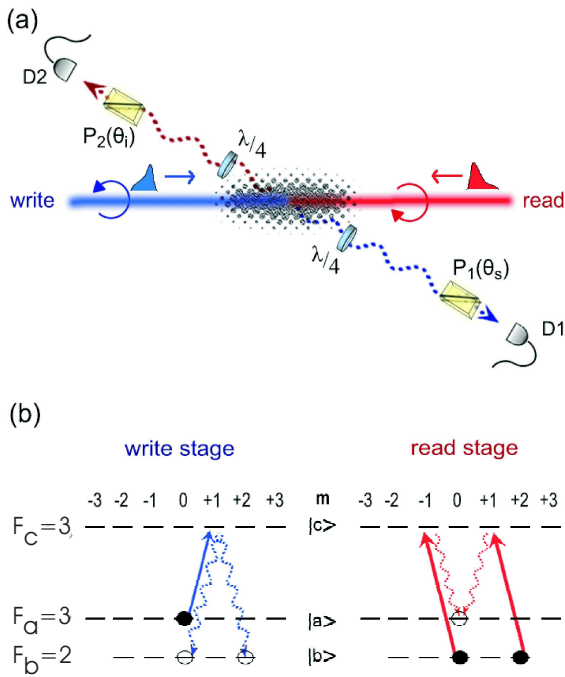,height=4.0in,width=3.1in}
\end{center}
\vspace{-1.2cm} \caption{ (a) Schematic of experimental setup. $P_1$ and $P_2$, polarizers; D1 and D2, detectors; $\lambda /4$,
quarter-waveplate. (b) The structure of atomic transitions leading to generation of atom-photon entanglement and of the subsequent read-out of
atomic qubit.}\label{TQ}
\end{figure}

The latter requirement is problematical as photons are difficult to store for an appreciable period of time. By contrast atomic qubits are long
lived and easily manipulated by laser fields, they are well suited for long term quantum information storage. Photonic qubits, however, can
propagate for relatively long distances in fibers without absorption, making them excellent carriers of quantum information. Entangled systems of
a single photon and a long-lived atomic qubit therefore offer an excellent building block for a quantum network.

A quantum repeater architecture can overcome the limitations of photons by inserting a quantum memory qubit into the quantum channel every
attenuation length or so \cite{briegel}. The idea is to generate entanglement between two neighboring atomic qubits, which can be done
efficiently since light will not be appreciably absorbed within the segment length. After entanglement between each pair of atomic qubits has
been established, a joint measurement on each neighboring pair of qubits is performed. The quantum states of all the intermediate qubits are
destroyed by the measurement, achieving entanglement swapping such that only the two atomic qubits at the two ends are entangled. These two
qubits can be used for QCKD, either with the Ekert protocol, that directly uses the entangled pair of qubits, or the BB84 protocol that performs
either remote state preparation or teleportation of a qubit \cite{ekert,bb84,bennett,zeilinger-t,boschi,kwiat1}. The rate of QCKD using a quantum
repeater protocol can scale polynomially with distance \cite{briegel}.

In the microwave domain, single Rydberg atoms and single photons have been entangled \cite{haroche}. An entangled state of an ion and a photon
has also been recently reported \cite{blinov,blinov1}. Cavity QED holds promise for generation of deterministic neutral atom-photon entanglement,
with single photon generation being an important step in that direction \cite{rempe,kimble}. Collective enhancement of atom-photon interactions
in optically thick atomic ensembles offers a somewhat simpler route towards this goal \cite{kuzmich,lukin,lukin3,zibrov,phillips,hau,johnsson}.
However, the atom-photon entanglement is often of a probabilistic character.

Duan, Lukin, Cirac, and Zoller (DLCZ) have developed a program of long-distance quantum networking based on atomic ensembles \cite{duan}. This
paradigm turns out to be remarkably similar to parametric down conversion, with the additional capability of atomic quantum memory. Non-classical
radiation has been produced from an atomic ensemble \cite{kuzmich2,eisaman,jiang,chou,eisaman1,polyakov}, as well as the preparation of a quantum
memory qubit based on two atomic ensembles with subsequent quantum state transfer onto a photonic qubit \cite{matsukevich}. These experiments
\cite{kuzmich2,eisaman,jiang,chou,eisaman1,polyakov,matsukevich} employed copropagating {\it write} and {\it read} laser fields and on-axis
Raman-scattered light was collected. In contrast to these works, Braje and coworkers pioneered off-axis four-wave mixing \cite{braje1} and
efficient photon-pair production \cite{braje2} in a cold atomic ensemble, using counter-propagating {\it write} and {\it read} fields deep in the
regime of electromagnetically-induced transparency.

In this Letter we report probabilistic entanglement of a collective atomic excitation and a photon (signal), achieved using the off-axis,
counter-propagating geometry of Braje {\it et al.} \cite{braje1,braje2}.  We propose and experimentally implement here an a qubit consisting of
two distinct mixed states of collective ground-state hyperfine coherence which contain one spin excitation. The entanglement of the signal photon
and the collective spin excitation is inferred by performing quantum state transfer of the atomic qubit onto a photonic qubit (idler)
\cite{matsukevich}, with one of the atomic states being converted into a right-hand polarized photon and the other into a left-hand polarized
one. Polarization correlations of the signal and the idler photons are subsequently recorded and found to be in violation of the Bell inequality.
The atom-photon entanglement is probabilistic, with the fundamental quantum state consisting mostly of vacuum. The entangled component of the
state is postselected by coincidence counting. This type of entanglement is similar to two-photon entanglement in spontaneous parametric
down-conversion (see \cite{mandel,zeilinger,mandel-wolf} and references therein), and to the ion-photon entanglement of Blinov {\it et al.}
\cite{blinov,blinov1}.

{\it Theory.} As illustrated in Fig.1(a), the right circularly polarized {\it write} pulse generates a cone of forward Raman scattering. We
collect a Gaussian mode centered around the momentum $\vec k_s$ that forms a $2^{\circ }$ angle with the {\it write} beam. Fig.1(b) indicates
schematically the structure of the three atomic levels involved, $|a\rangle ,|b\rangle $ and $|c\rangle $. The experimental sequence starts with
all of the atoms prepared in the unpolarized level $|a\rangle $. A {\it write} pulse tuned to the ${|a\rangle \rightarrow |c\rangle }$ transition
is directed into a sample of cold $^{85}$Rb atoms. The classical {\it write} pulse is so weak that less than one photon is scattered in this
manner on the ${|c\rangle \rightarrow |b\rangle }$ transition into the collected mode for each pulse.

Perturbation theory shows that the ensemble-photon density operator may be written as $|vac\rangle\langle vac|\otimes \rho_a+\epsilon \; \rho$
where $\rho$ has unit trace, and $\epsilon << 1$. Here $|vac\rangle$ is the photon vacuum state and $\rho_a$ the atomic ensemble vacuum state
density operator, corresponding to $N$ atoms each populating the Zeeman states $|a,m\rangle$ of level $|a\rangle$ with equal probability
$1/(2F_a+1)$. It is important to realize that the vacuum component in state $|vac\rangle\langle vac|\otimes \rho_a+\epsilon \; \rho$ has no
influence on the fidelity of DLCZ's quantum communication protocols due to built-in purification, even though $\epsilon \ll 1$ \cite{duan}.
Writing $|r\rangle$ and $|l\rangle$ as the normalized states of right and left circular polarization of the signal photon propagating towards the
detector in direction $\vec k_s$, we have that, in the ideal case
\begin{eqnarray} \nonumber
\rho =& \cos ^2 \eta |r\rangle \langle r| \hat s^{\dagger}_{-1} \rho_{a} \hat s_{-1} + \sin ^2 \eta|l\rangle \langle l| \hat s^{\dagger}_{1}
\rho_{a} \hat s_{1}
\\
&+ \cos \eta \; \sin \eta \left( |r\rangle \langle l| \hat s^{\dagger}_{-1} \rho_{a} \hat s_{1} + |l\rangle \langle r|\hat s^{\dagger}_{1}
\rho_{a} \hat s_{-1} \right )
\end{eqnarray}
where
\begin{equation}
\cos^2 \eta = \sum_{m} X_m^2 (-1)/[ \sum_{m} \sum_{\alpha=\pm1} X_m^2 (\alpha) ],
\end{equation}
with $m$ summed over $\{-F_a,...F_a\}$, and $X_m(\alpha) = C^{F_a,1,F_c}_{m,1,m+1} C^{F_c,1,F_{b}}_{m+1,\alpha,m+\alpha+1}$ is the product of the
relevant Clebsch-Gordan coefficients for the transition. The collective atomic spin excitation operators are given by
\begin{equation}
\hat s^{\dagger}_{\alpha} = \sum_{m} \left( \frac{X_m(\alpha)}{\sqrt{\sum_m X_m^2(\alpha)}} \right) \hat s^{\dagger}_{\alpha}(m)
\end{equation}
and
\begin{equation}
\hat s^{\dagger}_{\alpha}(m) = \sqrt{\frac{2F_a+1}{N}} \sum_{\mu =1}^{N} e^{-i \vec{ \Delta k_s} \cdot \vec r_{\mu} } | b, m+1+\alpha
\rangle_{\mu} \langle a,m|,
\end{equation}
where $\vec{ \Delta k_s} = \vec k_s - \vec k_w$, is the difference in the signal and {\it write} beam wave vectors and $\vec r_{\mu}$ is the
position of atom ${\mu}$. For weak states of excitation the collective spin operators satisfy bosonic commutation relations correct to $O(1/N)$:
$ [\hat s_{\alpha}(m),\hat s^{\dagger}_{\alpha'}(m')] = \delta_{\alpha,\alpha'} \delta_{m,m'}$ and $ [\hat s_{\alpha},\hat s^{\dagger}_{\alpha'}]
= \delta_{\alpha,\alpha'}$. Evaluating the coefficient $cos^2 \eta$ for the experimental conditions $F_a=F_c=3$, $F_{b}=2$, we find $\eta = 0.81
\times \pi/4$.

Detection of a photon by D1 produced by the $|c\rangle \rightarrow |b\rangle $ transition results in the sample of atoms containing, in the ideal
case, exactly one excitation in the related collective atomic mode. After a variable delay time $\Delta t$ (bounded by the lifetime of the
ground-state atomic coherences) we convert the atomic excitation into a single photon by illuminating the atomic ensemble with a pulse of light
near-resonant with the $|b\rangle \rightarrow |c\rangle $ transition and counter-propagating with respect to the {\it write} beam (Fig.1).  For
an optically thick atomic sample, the idler photon will be emitted with high probability into the mode determined by the phase-matching condition
$\vec k_i = \vec k_w +\vec k_r-\vec k_s$, with the atomic qubit state mapped onto a photonic one. Under the condition of collective enhancement
the atomic excitations generated by $\hat s_{\pm1}^\dagger $ map to orthogonal idler photon states up to a phase. Assuming equal mapping
efficiency, the number of correlated signal-idler counts registered by the detectors can be predicted on the basis of Eq.~(1). We find, by
carefully analyzing the measurement procedure,
\begin{eqnarray}
C\left(\theta _s,\theta _i\right) \propto \big[ (\cos\eta+\sin\eta )\cos\left(\theta _s - \theta _i\right)+ \nonumber \\
 (\cos\eta-\sin\eta)
\cos\left(\theta _s + \theta _i\right) \big]^2, \label{Fringes}
\end{eqnarray}
where $\theta_s$ and $\theta_i$ are the orientations of polarizers $P_1$ and $P_2$. Following Clauser-Horne-Shimony-Holt (CHSH)
\cite{chsh,walls}, we calculate the correlation function $E\left(\theta _s,\theta _i\right)$, given by
\begin{equation}
\frac{C\left(\theta _s,\theta _i\right)+C\left(\theta _s^\bot,\theta _i^\bot\right)-C\left(\theta _s^\bot,\theta _i\right)-C\left(\theta
_s,\theta _i^\bot\right)}{C\left(\theta _s,\theta _i\right)+C\left(\theta _s^\bot,\theta _i^\bot\right)+C\left(\theta _s^\bot,\theta
_i\right)+C\left(\theta _s,\theta _i^\bot\right)},
\end{equation}
where $\theta ^\bot = \theta + \pi /2$. The CHSH version of the Bell inequality is then $|S|\leq 2$ where
\begin{equation}
S=E\left(\theta _s,\theta _i\right) +
E\left({\theta_s}^\prime,\theta _i\right)+E\left(\theta _s,\theta
_i^\prime\right)-E\left(\theta _s^\prime,\theta _i^\prime\right).
\end{equation}
The maximum violation of the Bell inequality is achieved for a maximally entangled state with the canonical set of angles $\theta_s=-22.5^{\circ
}$, $\theta_i=0^{\circ }$, $\theta_s^\prime=22.5^{\circ }$ and $\theta_i^\prime=-45^{\circ }$: $S=2\sqrt{2}=2.83$. Based on the value $\eta =
0.81 \times \pi/4$ we find, ideally, $S=2.77$ which significantly violates the Bell inequality.

{\it Experiment.}  A magneto-optical trap (MOT) of $^{85}$Rb is used to provide an optically thick atomic cloud for our experiment (Fig.1). The
ground levels $\{|a\rangle;|b\rangle \}$ correspond to the $5S_{1/2},F_{a,b}=\{3,2\}$ levels, while the excited level $|c\rangle$ represents the
$\{5P_{1/2},F_c=3\}$ level of the $D_1$ line at 795 nm. The experimental sequence starts with all of the atoms prepared in level $|a\rangle $.
The ``dark" period lasts 640 ns, with the whole cycle taking 1.5 $\mu$s. All the light responsible for trapping and cooling is shut off during
the dark period, with the trapping light shut off about 200 ns before the repumping light to empty the $F=2$ hyperfine level. The quadrupole
magnetic field of the MOT is switched off for the duration of the measurement sequence. The ambient magnetic field is compensated by three pairs
of Helmholtz coils.

A 130 ns long  {\it write} pulse tuned to the ${|a\rangle \rightarrow |c\rangle }$ transition is focused into the MOT with a Gaussian waist of
about $400$ $\mu$m. The light induces spontaneous Raman scattering via the ${|c\rangle \rightarrow |b\rangle }$ transition. The scattered light
goes through the quarter-wave plate to map circular polarizations into linear ones, then passes through polarizer $P_1$ (set at angle $\theta
_s$) and impinges onto a single photon detector D1.

After a user-programmable delay $\Delta t$, a 120 ns long {\it read} pulse, with circular polarization opposite to that of the {\it write} pulse,
tuned to the ${|b\rangle \rightarrow |c\rangle }$ transition illuminates the atomic ensemble. This accomplishes a transfer of the memory state
onto the single photon (idler) emitted by the ${|c\rangle \rightarrow |a\rangle }$ transition. After passing through the quarter-wave plate and
polarizer $P_2$ set at angle $\theta _i$, the idler photon is directed onto a single-photon detector D2.

\begin{figure}[htp]
\begin{center}
\leavevmode  \psfig{file=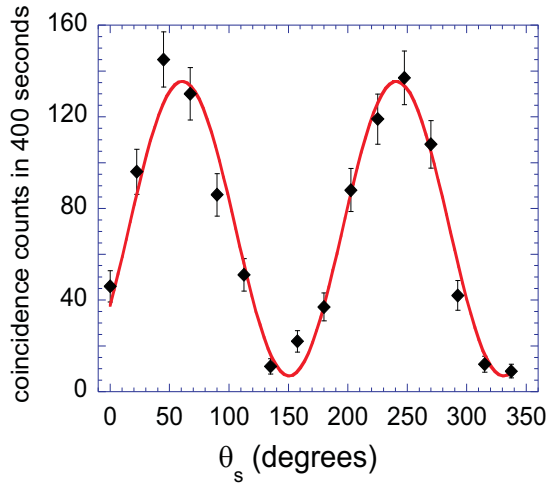,height=2.5in,width=2.8in}
\end{center}
\vspace{-0.6cm} \caption{ Measured coincidence fringe for $\theta _i =67.5^{\circ }$. The curve is a fit based on Eq.(5), augmented by a
background contribution, with $\eta = 0.81 \times \pi/4$, with visibility and amplitude being adjustable parameters. The visibility of the fit is
$90\%$. Uncertainties are based on the statistics of the photon counting events.}\label{TQ}
\end{figure}
\begin{table}
\caption{\label{tab:table1} Measured correlation function $E(\theta _s, \theta _i)$ and $S$ for  $\Delta t = 200$ ns delay between {\it write}
and {\it read} pulses; all the errors are based on the statistics of the photon counting events.}
\begin{ruledtabular}
\begin{tabular}{ccccc}
$\theta _s$ & $\theta _i $& $E(\theta_s, \theta _i)$  \\
\hline
-22.5  & 0     &  $0.641  \pm 0.024$   \\
-22.5  & -45 &  $0.471  \pm 0.029$   \\
 22.5  & 0     &  $0.587  \pm 0.027$   \\
  22.5  & -45 &  $-0.595 \pm 0.027$   \\
     &    & $S=2.29 \pm 0.05$           \\
\end{tabular}
\end{ruledtabular}
\end{table}

Both {\it write/read} and signal/idler pairs of fields  are counter-propagating. The waist of the signal-idler mode in the MOT is about 150
$\mu$m. The four-wave mixing signal is used to align the single mode fibers collecting signal and idler photons, and to optimize the overlap
between the pump and probe modes \cite{braje1}. The value of delay $\Delta t$ between the application of the {\it write} and {\it read} pulses is
200 ns. The electronic pulses from the detectors are gated with 140 ns and 130 ns windows centered on the time determined by the {\it write} and
{\it read} light pulses, respectively. Afterwards, the electronic pulses are fed into a time-interval analyzer (with 2 ns time resolution). In
order to measure the correlation between the photons produced by the {\it write} and {\it read} pulses, the output of D1 is fed into the ``Start"
input of a time-interval analyzer, and the output of D2 is fed into the ``Stop" input.

\begin{figure}[btp]
\begin{center}
\leavevmode  \psfig{file=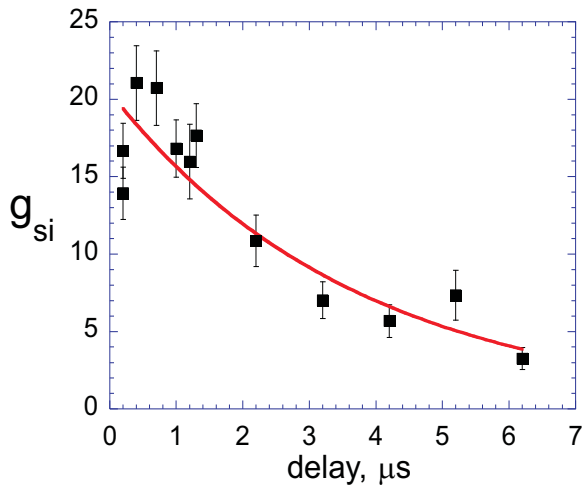,height=2.5in,width=3.0in}
\end{center}
\vspace{-0.5cm} \caption{Normalized signal-idler intensity correlation function $g_{si}$ as a function of storage time. Uncertainties are based
on the statistics of the photon counting events. The full curve is the best exponential fit with time constant $\tau = 3.7 \mu$s.} \label{TQ}
\end{figure}

A typical interference fringe in the signal-idler coincidence detection is displayed in Fig.2. In order to infer probabilistic atom-photon
entanglement, we calculate the degree of Bell inequality violation $|S|\leq 2$ \cite{chsh,walls}. Table 1 presents measured values for the
correlation function $E\left({\theta_s},\theta _i\right)$ using the canonical set of angles $\theta_s,\theta _i$. We find $S=2.29 \pm 0.05 \nleq
2$ - a clear violation of the Bell inequality. The value of $S$ is smaller than the ideal value of $2.77$ due to experimental imperfections,
particularly non-zero counts in the minima of interference curves that arise as the result of the finite value of the normalized signal-idler
intensity correlation function $g_{si}$ \cite{mandel-wolf,kuzmich2,chou} shown in Fig. 3. To our knowledge, this is the first observed violation
of the Bell inequality involving a collective excitation.

The effective detection efficiencies as determined by the ratios of the coincidence signal-idler count rate $R_{si}$ to singles count rates $R_s$
and $R_i$ are $\alpha _{s,i} = R_{si}/R_{i,s} \simeq 0.02$. In all cold atomic ensemble experiments within the DLCZ program reported to date, the
quadrupole magnetic field of the MOT has been the main source of the atomic memory decoherence (limiting storage times on the order of 100 ns
\cite{kuzmich2,chou,polyakov,matsukevich}). In this work, we have switched off the quadrupole field for the duration of our protocol, and the
coherence time has increased to several $\mu$s, as is evident from the measured normalized intensity correlation function $g_{si}$ displayed in
Fig.3 (the length of the dark period was increased up to 7 $\mu$s for this measurement at the expense of lower count rate).

The robustness and relative simplicity of probabilistic atom-photon entanglement hold promise for the realization of a distributed quantum
network involving the interconnection of several similar elements. We are currently investigating connecting two such quantum nodes.

We gratefully acknowledge illuminating discussions with D. A. Braje, M. S. Chapman, and S. E. Harris. We thank  E.T. Neumann for experimental
assistance. This work was supported by NASA, National Science Foundation, Office of Naval Research Young Investigator Program, Research
Corporation, Alfred P. Sloan Foundation, and Cullen-Peck Chair.

\end{document}